# Modelling and Control of a Buck Converter Using State-Space Averaging and Classical Feedback Techniques

Sampson E. Nwachukwu, *Graduate Student Member, IEEE*

*Abstract*—This study presents the modeling, control design, and performance analysis of a DC-DC buck converter using state-space averaging techniques. Buck converters are essential in modern power electronics for regulating DC voltages in renewable energy and electric vehicle systems. The paper first introduces the basic operation of buck converters and emphasizes the need for voltage regulation through closed-loop control systems. A state-space averaged model is derived to simplify the nonlinear switched dynamics, enabling a more effective analysis and controller design. The small-signal transfer function from the duty cycle to the output voltage is obtained to support control development. In addition, the Proportional-Integral (PI) control based on the frequency-domain method was explored. The PI controller was tuned to achieve various phase margins and is evaluated through Bode plots, step responses, and performance metrics, revealing trade-offs between overshoot, settling time, and steady-state error. A complete simulation of the controlled buck converter verifies its ability to maintain a stable output voltage across wide input voltage variations. The results validate the effectiveness of state-space averaging in control design and highlight the robustness of feedback systems in power electronic converters.

## I. INTRODUCTION

The primary factors in recent years for obtaining green energy and using renewable power sources, including photovoltaic (PV), fuel cells, and wind generation, have been the depletion of fossil fuels and environmental concerns. To provide a regulated output voltage from an unadjusted voltage resource, a DC-DC converter is necessary in the majority of systems [1]. Simply put, a DC-DC converter is a device that converts a DC source to a DC source. With a continual turn ratio, a DC converter may be seen as the DC version of an AC transformer. It may be used to step up or step down a DC voltage source, just like a transformer [2]. Almost every electronic gadget in use today has DC-DC converters, as all semiconductors are powered by DC sources [1]. For example, trolley cars, forklift trucks, electronic vehicles, and mine haulers employ DC-DC converters for traction motor control. DC-DC converters also offer quick dynamic response, high efficiency, and smooth acceleration control. In order to provide a DC source, particularly for the current source inverter, DC converters are utilized in DC voltage regulators as well as in combination with an inductor [2].

It is possible to employ the DC converters as switching mode regulators to change an unregulated DC voltage into a regulated DC output voltage. Typically, an IGBT, MOSFET, or BJT is used as the switching driver, and a pulse width modulator (PWM) at a set frequency is used to perform the control. The voltage controller may be linear, switch-mode, inductor-based, or a switch-capacitor charge pump. Every voltage controller has pros and cons, but the optimal technique depends on the needs of the specific application. Four switching-mode converter configurations exist, including Cuk, boost, buck, and buck-boost converters [2]. Nonetheless, many of these converters are subjected to comparable analysis and control techniques [3]. This paper presents the procedure for controlling the voltage of a buck converter to maintain a constant output voltage.

The remainder of this study is organized as follows: Section II presents the principles of operation of a buck converter. Section III presents the buck converter voltage control. Section IV presents the mathematical formulation of a buck converter. Section V presents the control design of a buck converter. Section VI presents the simulation results. Finally, Section VII presents the conclusion of the study and recommendations for future study.

## II. PRINCIPLES OF OPERATION OF A BUCK CONVERTER

Typically, the average output voltage, $V_o$, of a buck converter is lower than the input voltage, $V_g$. As shown in Fig. 1, an inductor and two switches—typically a diode and a transistor switch—for regulating the inductor make up a buck converter, which is a step-down DC-DC converter. The inductor's energy is stored by varying the connection of the induction to the source voltage, and the energy is subsequently released to the load [2]. Reverse-biasing of the diode occurs when the controller switches on the transistor because $V_B = -V_g$. To prevent the diode from becoming forward-biased, $V_g$ must be positive. The inductor current, $I_L$ is conducted by the transistor. In order for the transistor to conduct in a forward direction, this current must also be positive [2], [4].

For $I_L$ to continue flowing after the controller switches off the transistor, the diode must also switch on. A reduction in $I_L$ occurs when the transistor is turned off. The diode switches on when the inductor voltage drops enough to forward-bias it because $V_L = L\frac{dI_L}{dt}$. This type of operation is commonly referred to as "freewheeling diodes". Given that $I_B = I_L$, the diode cannot be forward-biased unless $I_L$ is positive. To prevent the transistor from operating in reverse blocking mode, the voltage $V_g$, which the transistor blocks, must be positive [2], [4].

As shown in Fig. 2, the transistor in the buck converter has a duty ratio $D$ and runs at a constant switching period, $T_s$ and frequency, $f_s$. Here, the output capacitor ripple voltage is regarded as insignificant, the inductor current is believed to be continuous, and the circuit components are lossless. The negative and positive inductor volt-seconds in a switching cycle are equated to determine the connection between the steady-state output voltage and the duty ratio [2]–[4].

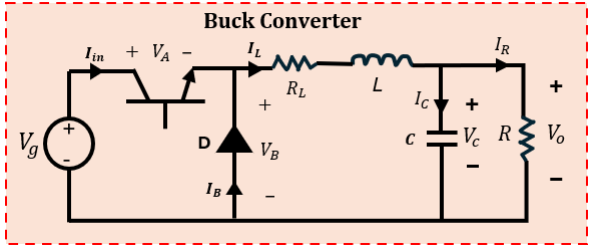

Fig. 1. Schematic of a buck converter

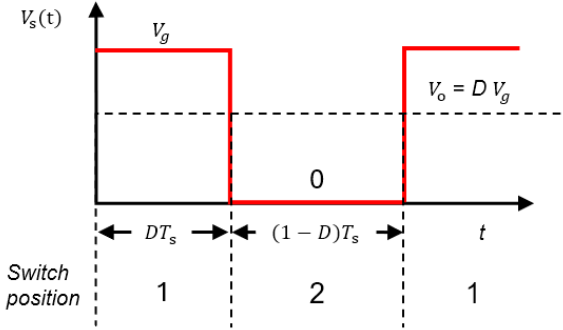

Fig. 2. Switch changes of a buck converter voltage level

When operating in a steady condition, the volt-seconds must balance, as expressed in (1). Also, $V_o$ needs to be lower than or equal to $V_g$ since D is between 0 and 1 [2]–[4].

$$(V_g - V_o)DT_s = V(1 - D)\,T_s \Rightarrow V_o = DV_g \qquad (1)$$

III. BUCK CONVERTER VOLTAGE CONTROL

In all cases, a feedback controller, shown in Fig. 3, is necessary for converter systems. As shown in the figure, the output voltage, $V_o$, in a typical DC-DC converter application needs to remain constant despite variations in the load resistance, R, or the input voltage $V_g$. To achieve this, a circuit that modifies the duty cycle, d, or converter control input, is developed so that the output voltage is controlled to match the required reference voltage, $V_{ref}$ [4]. Similarly, the feedback system should be designed so that the output voltage is precisely controlled and unaffected by changes in the load current, $I_R$. Furthermore, the feedback system must be stable and have characteristics that satisfy requirements, including settling time, transient overshoot, and steady-state control [4].

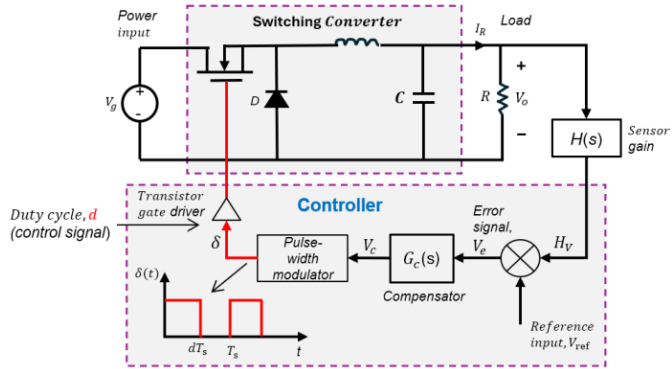

Fig. 3. Schematic of a buck converter voltage control [4]

In order to develop the system shown in Fig. 3, a dynamic model of the switching converter is required. This involves determining the output voltage change in response to changes in the duty cycle, input voltage, and load current, and the transfer functions of the system's small signal.

IV. MATHEMATICAL FORMULATION OF A BUCK CONVERTER

As discussed earlier, a converter control system must be able to maintain a consistent output voltage regardless of changes in the load current, $I_R$ and DC source voltage, $V_g$. Also, $V_o$ has no effect on load conditions based on the output voltage's steady state in (1). Nonetheless, load variations have a transient impact on the output voltage, potentially leading to notable variations from the level of the steady state. Additionally, circuit losses in a real system provide an output voltage dependence on the steady-state load current, which the control system must adjust for [2]–[4]. Therefore, the following sections present the dynamic modeling of the buck converter using approximate averaging modeling.

A. Dynamic Modelling of Buck Converter Using State-Space Average Modelling

The components of power electronic converters are coupled in a variety of configurations that change frequently due to their intrinsic switching behavior; each configuration is defined by a different set of equations. Consequently, the transient analysis and control design of converters are challenging due to the sequential solution of many equations. This problem can be resolved by using the averaging approach. By simply calculating a linearly weighted average of the many equations for each converter's switched configuration, a single equation that roughly describes the converter throughout a switching cycle may be developed. Thus, the buck converter is modeled here using state-space averaging, which is the most often used averaging method [3]. The buck converter circuit's mode of operation that corresponds to the transistor states in Fig. 1 is presented in the following subsections.

i) Mode 1

Fig. 4 illustrates the second operating mode of the converter circuit. Here, the transistor switch is ON, i.e., $q(t) = 1$.

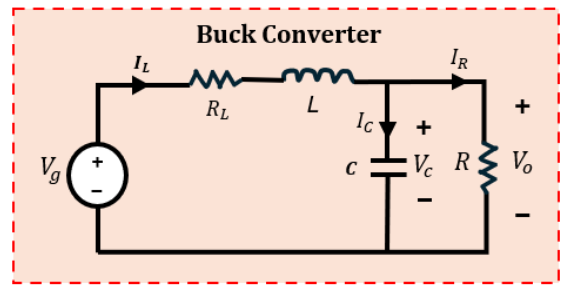

Fig. 4. Schematic of a buck converter when $q(t) = 1$

Applying KVL in loop 1 of Fig. 4 above, we have:

$$V_g = I_L R_L + L\frac{dI_L}{dt} + I_R R \qquad (2)$$

Applying KCL at the node, we have:

$$I_R = I_L - I_C \quad (3)$$

Applying KVL in loop 2, we have:

$$V_C = I_R R \quad (4)$$

But, $I_C = C\frac{dV_C}{dt}$

Thus,

$$I_R = I_L - C\frac{dV_C}{dt} \quad (5)$$

From (4),

$$V_C = R\left(I_L - C\frac{dV_C}{dt}\right) = RI_L - RC\frac{dV_C}{dt} \quad (6)$$

Therefore,

$$\frac{dV_C}{dt} = \frac{I_L}{C} - \frac{V_C}{RC} \quad (7)$$

Substituting (7) into (5), we have

$$I_R = I_L - C\left(\frac{I_L}{C} - \frac{V_C}{RC}\right) = \frac{V_C}{R} \quad (8)$$

Then, from (1),

$$L\frac{dI_L}{dt} = V_g - I_L R_L - R\left(\frac{V_C}{R}\right) = V_g - I_L R_L - V_C \quad (9)$$

So,

$$\frac{dI_L}{dt} = \frac{V_g}{L} - \frac{I_L R_L}{L} - \frac{V_C}{L} \quad (10)$$

Also, the output voltage is:

$$V_o = V_C \quad (11)$$

Thus, expressing the derived system model in a standard state space model, we have:

$$\begin{bmatrix} I'_L \\ V'_C \end{bmatrix} = \begin{bmatrix} -\frac{R_L}{L} & -\frac{1}{L} \\ \frac{1}{C} & -\frac{1}{RC} \end{bmatrix}\begin{bmatrix} I_L \\ V_C \end{bmatrix} + \begin{bmatrix} \frac{1}{L} \\ 0 \end{bmatrix} V_g \quad (12)$$

In (12), the input vector, $u$ has components $V_g$, whereas the state vector $x$ consists of the inductor current, $I_L$ and output capacitor voltage, $V_C$. Thus, the output voltage is expressed as:

$$V_o = y = \begin{bmatrix} 0 & R \end{bmatrix}\begin{bmatrix} I_L \\ V_C \end{bmatrix} \quad (13)$$

### 2) Mode 2

Fig. 5 illustrates the second operating mode of the converter circuit. Here, the transistor switch is OFF, i.e., $q(t) = 0$.

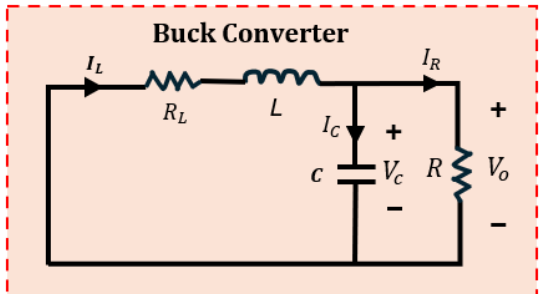

Fig. 5. Schematic of a buck converter when $q(t) = 0$

Since the switch is off, $V_g = 0$, thus the state-space model becomes:

$$\begin{bmatrix} I'_L \\ V'_C \end{bmatrix} = \begin{bmatrix} -\frac{R_L}{L} & -\frac{1}{L} \\ \frac{1}{C} & -\frac{1}{RC} \end{bmatrix}\begin{bmatrix} I_L \\ V_C \end{bmatrix} + \begin{bmatrix} 0 \\ 0 \end{bmatrix} V_g \quad (14)$$

$$V_o = y = \begin{bmatrix} 0 & R \end{bmatrix}\begin{bmatrix} I_L \\ V_C \end{bmatrix} \quad (15)$$

In this study, we set $V_g = 30$ V, $V_o = 15$ V, $R = 10\ \Omega$, $I_o = 1.5$ A, $L = 250\ \mu H$, $R_L = 0.2\ \Omega$, and $C = 30$ mF. Thus,

$$A_1 = A_2 = \begin{bmatrix} -\frac{R_L}{L} & -\frac{1}{L} \\ \frac{1}{C} & -\frac{1}{RC} \end{bmatrix} = \begin{bmatrix} -800 & -4000 \\ 33.333 & -3.333 \end{bmatrix};$$

$$B_1 = \begin{bmatrix} \frac{1}{L} \\ 0 \end{bmatrix} = \begin{bmatrix} 4000 \\ 0 \end{bmatrix};\ B_2 = \begin{bmatrix} 0 \\ 0 \end{bmatrix};\text{ and}$$

$$C_1 = C_2 = \begin{bmatrix} 0 & R \end{bmatrix} = \begin{bmatrix} 0 & 10 \end{bmatrix}$$

### 3) Determining the Equilibrium Point Around the Duty Cycle and Input Voltage

A weighted average of the above equations is used to develop the converter's state-space averaged model. Since $x = \begin{bmatrix} I_L \\ V_C \end{bmatrix}$, the combined state-space model can be determined as:

$$\dot{x}(t) = q(A_1 x + B_1 V_g) + (1 - q)(A_2 x + B_2 V_g) \quad (16)$$

$$V_o = qC_1 x + (1 - q)C_2 x \quad (17)$$

Applying the state space averaging approach, we have:

$$\dot{x}(t) = d(t)(A_1 x(t) + B_1 V_g) + (1 - d(t))(A_2 x(t) + B_2 V_g) \quad (18)$$

Thus,

$$V_o = d(t)C_1 x(t) + (1 - d(t))C_2 x(t) \quad (19)$$

Since $d(t) = D$, equation (18) becomes:

$$D(A_1 x_o + B_1 V_g) + (1 - D)(A_2 x_o + B_2 V_g) = 0 \quad (20)$$

Then, the equilibrium points around the nominal input voltage, $V_g$ and duty cycle, $D$ becomes:

$$x_o = -(DA_1 + (1 - D)A_2)^{-1}(DB_1 + (1 - D)B_2)V_g \quad (21)$$

Since $A_1 = A_2$,

$$x_o = -(A_1)^{-1} D B_1 V_g \quad (22)$$

Therefore,

$$(A_1)^{-1} = \frac{1}{|A_1|}\begin{bmatrix} -800 & -4000 \\ 33.333 & -3.333 \end{bmatrix}^{-1}$$

$$|A_1| = (2{,}666.4 + 133{,}332) = 135{,}998.4$$

$$(A_1)^{-1} = \frac{1}{135{,}998.4}\begin{bmatrix} -3.333 & 4000 \\ -33.333 & -800 \end{bmatrix}$$

$$DB_1 V_g = \begin{bmatrix} 120{,}000D \\ 0 \end{bmatrix}$$

$$x_o = -\frac{1}{135{,}998.4}\begin{bmatrix} -3.333 & 4000 \\ -33.333 & -800 \end{bmatrix}\begin{bmatrix} 120{,}000D \\ 0 \end{bmatrix}$$

$$= -\frac{1}{135{,}998.4}\begin{bmatrix}-399{,}960D\\-3{,}999{,}960D\end{bmatrix}=\begin{bmatrix}3D\\29.4D\end{bmatrix}$$

Therefore,

$$x_o = \begin{bmatrix}I_L\\V_C\end{bmatrix} = \begin{bmatrix}3D\\29.4D\end{bmatrix}$$

Thus,

$$V_o = V_C = 29.4D$$
$$15 = 29.4D$$

Therefore, $D = 0.51$, and the equilibrium points becomes:

$$x_o = \begin{bmatrix}I_L\\V_C\end{bmatrix} = \begin{bmatrix}1.51\\15\end{bmatrix}$$

Although the averaging method has eliminated all information on the variables' switching frequency ripple component, equation (16) roughly depicts the operation of the converter over a large number of cycles. Two major requirements must be met for the averaging approximation to be valid: first, the state variables in the two circuit configurations must evolve in a roughly linear fashion. Second, compared to the average component, the state variables' switching frequency ripple component needs to be minimal. Simple DC-DC converters often meet both of these requirements [3].

However, the duty ratio ($D$), the converter's control input, is not an element in the input vector but rather occurs in the $B$ matrix of the averaged model in (16). Thus, the averaged model is time-variable and challenging to solve. Therefore, equation (16) is linearised by taking into account slight fluctuations in the variables in order to simplify the model. Also, a steady state and a small signal are added together to represent each variable [3]. So, the local linearization if $V_g$ is not fixed becomes:

$$\tilde{x}(t) = x(t) - x_o,\ \tilde{d}(t) = x(t) - D,\ \tilde{V}_g(t) = V_g(t) - V_g(t),\ \tilde{y}(t) = v(t) - V_o$$

After substituting the above variables in (16), the equation is multiplied out, and the products of small-signal values are ignored. The linear equation that follows, which links minor changes in the variables, is obtained by subtracting the state-space of the variables [3]. Then, the Jacobian matrix with respect to $x$ and $d$ becomes:

$$\frac{d\tilde{x}(t)}{dt} = (DA_1 + (1-D)A_2)\tilde{x}(t) + [(A_1 - A_2)x_o + (B_1 - B_2)V_g]\tilde{d}(t) = A\tilde{x}(t) + B\tilde{d}(t) \quad (23)$$

Thus,

$$A = (DA_1 + (1-D)A_2) = A_1 = A_2 = \begin{bmatrix}-800 & -4000\\33.333 & -3.333\end{bmatrix}$$

$$B = (A_1 - A_2)x_o + (B_1 - B_2)V_g = \begin{bmatrix}4000\\0\end{bmatrix}V_g = \begin{bmatrix}120{,}000\\0\end{bmatrix}$$

Also, the voltage control is:

$$\tilde{y} = \tilde{v} = C = \begin{bmatrix}0 & 1\end{bmatrix}\tilde{x}$$

The transfer function from the duty cycle, $d(t)$, to $V_o$ becomes:

$$\frac{\tilde{y}(s)}{D(s)} = C(SI - A)^{-1}B$$

$$\frac{\tilde{y}(s)}{D(s)} = \begin{bmatrix}0 & 1\end{bmatrix}\begin{bmatrix}s+800 & 4000\\-33.333 & s+3.333\end{bmatrix}^{-1}\begin{bmatrix}120{,}000\\0\end{bmatrix}$$

$$|SI - A| = \begin{bmatrix}s+800 & 4000\\-33.333 & s+3.333\end{bmatrix}$$

$$= S^2 + 803.333s + 135{,}998.4$$

$$(SI - A)^{-1} = \begin{bmatrix}s+3.333 & -4000\\33.333 & s+800\end{bmatrix}$$

Thus,

$$\frac{\tilde{y}(s)}{D(s)} = \begin{bmatrix}0 & 1\end{bmatrix}\frac{1}{S^2 + 803.333s + 135{,}998.4}$$
$$\times \begin{bmatrix}s+3.333 & -4000\\33.333 & s+800\end{bmatrix}\begin{bmatrix}120{,}000\\0\end{bmatrix}$$

$$\frac{\tilde{y}(s)}{D(s)} = \frac{1}{S^2 + 803.333s + 135{,}998.4}\begin{bmatrix}0 & 1\end{bmatrix}$$
$$\times \begin{bmatrix}120{,}000 \times (s+3.333)\\120{,}000 \times (33.333)\end{bmatrix}$$

$$\frac{\tilde{y}(s)}{D(s)} = \frac{3{,}999{,}960}{S^2 + 803.333s + 135{,}998.4}$$

## V. CONTROL DESIGN OF BUCK CONVERTER

### A. Unit-Step Response of the System

A real-time variation in the reference input is represented by the step-function input. A step input, for instance, indicates the abrupt rotation of a mechanical shaft if the input is the shaft's angular position. Fig. 6 illustrates the step function as a function of time.

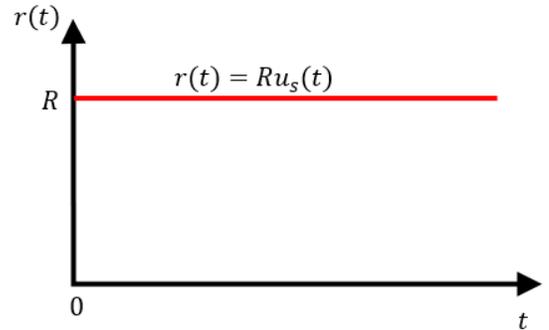

Fig. 6. Unit-step function signal

The step function or magnitude $R$ can be represented mathematically as [5]:

$$r(t) = R \quad t \geq 0$$
$$= 0 \quad t < 0 \quad (24)$$

where $R$ is the real constant. Or,

$$r(t) = Ru_s(t) \quad (25)$$

where $u_s(t)$ is the unit-step function.

The step function's first immediate amplitude leap provides valuable information on how quickly a system responds to inputs with sudden changes, making it an excellent test signal. Additionally, because of the jump discontinuity, the step function is equal to applying many sinusoidal signals with a large range of frequencies, since it encompasses a broad spectrum of frequencies in concept [5]. Additionally, a typical linear control system's unit-step response is shown in Fig. 7. Here, when describing linear control systems in the time domain, the following performance criteria are frequently applied with regard to the unit-step response [5]:

- **Delay time:** When the step response takes 50% of its final value, it is said to have a delay time, $t_d$.

- **Settling time, $t_s$:** The time needed for the step reaction to diminish and remain within a given percentage of its peak value. Here, 5% is an often used number.

- **Rise time, $t_r$:** The amount of time needed for the step response to increase from 10% to 90% of its peak value.

- **Steady-state error:** The difference between the reference input and the output at the point of steady state (t → ∞).

- **Maximum overshoot:** Expressed as a percentage of the step response's peak value, i.e,

$$\% \max.\text{overshoot} = \frac{maximum\ overshoot}{y_{ss}} \times 100\% \quad (26)$$

where,

$$maximum\ overshoot = y_{max} - y_{ss} \quad (27)$$

where $y_{ss}$ is the unit step response, $y(t)$ steady-state value. $y_{max}$ is the unit step response's maximum value.

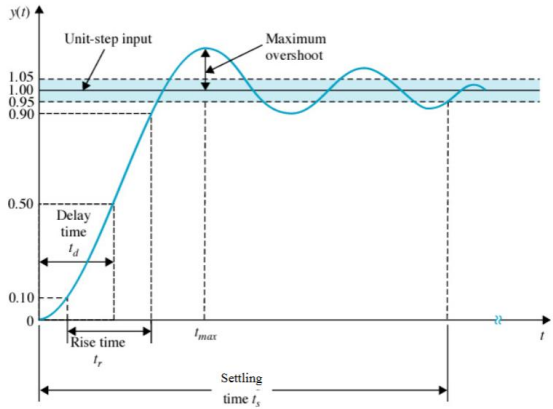

Fig. 7. Time domain description of a control system's unit-step function [5]

### B. Proportional Integral (PI) Controller Design Using Frequency Domain Method

There are no standard techniques for presenting a system design that satisfies time-domain performance requirements, including rising time, delay time, settling time, maximum overshoot, etc. In contrast, an extensive range of graphical techniques that are not exclusive to low-order systems are accessible in the frequency domain. Given that a linear system's time-domain and frequency-domain performances have correlated relations, it is crucial to understand that the frequency-domain features may be used to anticipate the system's time-domain qualities [5].

#### 1) Frequency Response of Closed-Loop Systems

The closed-loop transfer function for the single-loop control system is expressed as:

$$M(s) = \frac{Y(s)}{R(s)} = \frac{G(s)}{1+G(s)H(s)} \quad (28)$$

Substituting $s$ with $j\omega$, the equation for sinusoidal steady-state analysis becomes:

$$M(j\omega) = \frac{Y(j\omega)}{R(j\omega)} = \frac{G(j\omega)}{1+G(j\omega)H(j\omega)} \quad (29)$$

In terms of its magnitude and phase, the sinusoidal steady-state transfer function $M(j\omega)$ may be defined as follows:

$$M(j\omega) = |M(j\omega)| < M(j\omega) \quad (30)$$

Then, the magnitude of $M(j\omega)$ becomes:

$$|M(j\omega)| = \left|\frac{G(j\omega)}{1+G(j\omega)H(j\omega)}\right| \quad (31)$$

The phase of $M(j\omega)$ is:

$$< M(j\omega) = \emptyset_M(j\omega) = < G(j\omega) - < G(j\omega)H(j\omega) \quad (32)$$

If $M(s)$ is an electric filter's input-output transfer function, then the input signal's filtering properties can be determined by the magnitude and phase of $M(j\omega)$. A control system's typical gain and phase characteristics are shown in Fig. 8. In the figure, the highest possible value of $|M(j\omega)|$ is the resonant peak $M_r$. When the peak resonance $M_r$ takes place the frequency is known as the resonant frequency $\omega_r$. When $|M(j\omega)|$ falls to 70.7% of its zero-frequency value or 3 dB below it, that frequency is known as the bandwidth, BW [5].

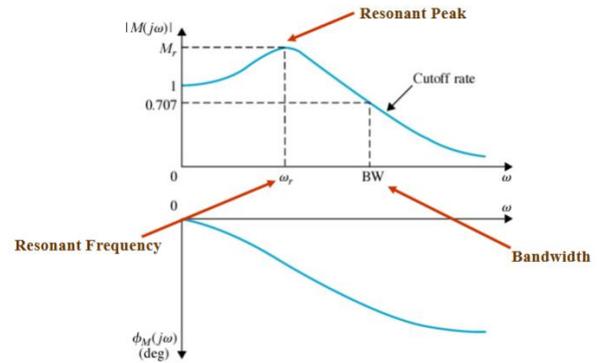

Fig. 8. Feedback control system gain-phase characteristics [5].

#### 2) Stability Analysis with Bode Plot

A transfer function's Bode plot is a vital graphical tool for frequency-domain linear control system analysis and design. Because the magnitude and phase curves may be approximated from their asymptotic qualities without the need for precise drawing, Bode plots were sometimes referred to as "asymptotic plots" prior to the invention of computers. The Bode plot makes it easier to determine gain margin, phase margin, phase crossover, and gain crossover than other approaches, such as the Nyquist plot. Also, the Bode plot makes it easier for designers to see the impact of adding controllers and their settings [5]. A diagrammatic representation of the Bode plot for determining phase margin and gain margin is shown in Fig. 9. As shown in the figure, if the amplitude of $L(j\omega)$ at the phase crossover is negative in dB, the system is stable, and the gain margin is positive. The gain margin is therefore measured below the 0 dB-axis. A negative gain margin indicates instability in the system if it is measured above the 0 dB-axis. The system is stable if the phase of $L(j\omega)$ is higher than $-180°$ at the gain crossover, and the phase margin is positive. In other words, the phase margin is determined above $-180°$ axis. The system is unstable and the phase margin is negative if it is determined below the $-180°$ axis [5].

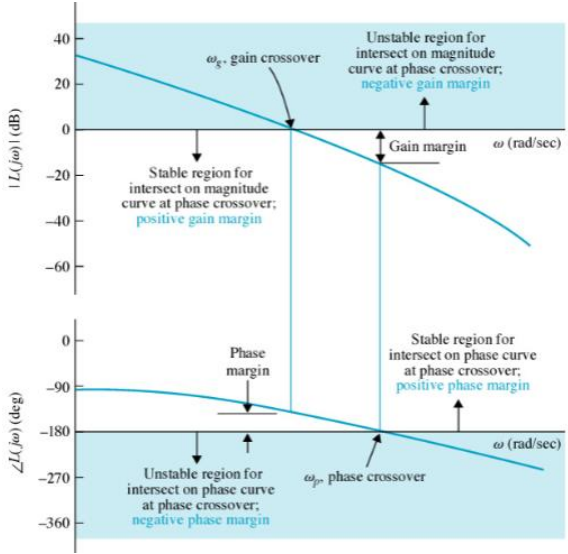

Fig. 9. A plot showing the estimation of the phase margin and gain margin on the Bode plot [5].

*3) PI Controller Design Based on Frequency Domain*

Traditionally, a simple amplifier with a fixed gain, $K$ serves as the controller. Since a proportional constant connects the control signal at the controller's output to its input, this kind of control action is properly called proportional-integral (PI) control. The PI controller's transfer function for frequency-domain design is expressed as:

$$G_c(s) = K_P + \frac{K_I}{s} = \frac{K_I\left(1 + \frac{K_P}{K_I}\right)}{s} \quad (33)$$

where $K_P$ is a positive gain that ranges from 0 to ∞, and the step input's steady-state errors can be eliminated using $K_I$.

To achieve a certain phase margin, the frequency-domain design process for the PI control is described as follows [5]:
- The loop gain is adjusted in accordance with the steady-state output condition to create the Bode plot of the system's uncompensated forward-path transfer function $G_o(s)$.
- From the Bode plot, the uncompensated system's phase margin and gain margin are calculated.
- At the new gain-crossover frequency $\omega'_g$, the PI controller must supply attenuation equal to the magnitude curve's gain in order to reduce the uncompensated transfer function's magnitude curve to 0 dB.

The compensated system often has a longer settling time and a slower rising time because the PI controller functions effectively as a low-pass filter. Choosing the zero at $s = -K_I/K_P$ to be somewhat close to the origin and away from the most important poles of the process is a workable technique for developing the PI control. However, $K_P$ and $K_I$ values need to be considerably small. A typical feedback control system with a PI controller is shown in Fig. 10.

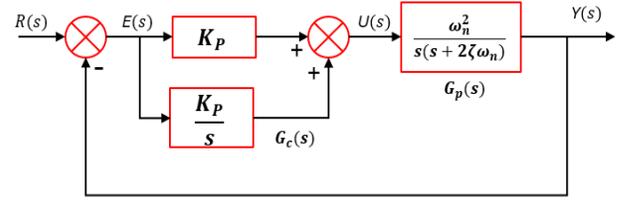

Fig. 10. Feedback control system with PI controller.

## VI. SIMULATION RESULTS

In this section, the closed-loop system is simulated to analyze its performance based on the unit-step response, root locus method, and frequency method.

*A. Unit-Step Response Simulation Results*

Here, the closed-loop system is simulated to analyze its performance under the unit-step response. The control system with the buck converter transfer function shown in Fig. 11 is modeled on MATLAB, whereas the output waveform and the response of the system are shown in Fig. 12. As shown in Fig. 12, the system responds quickly to the input. However, it has a high overshoot oscillation of the amplitude signal and settles near 1, at about 0.01 s, with minor steady-state error.

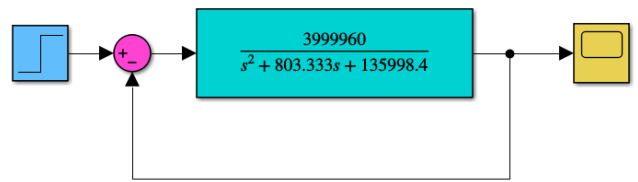

Fig. 11. Modeled feedback control system for unit step response.

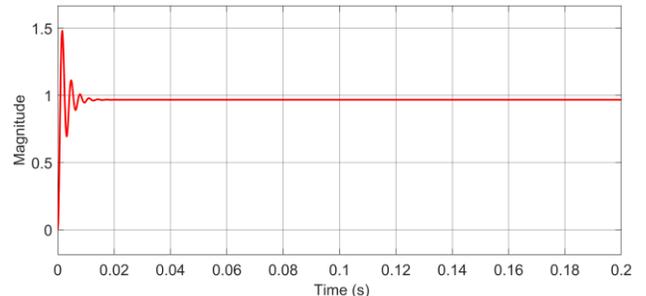

Fig. 12. Simulation result for unit step response.

### A. Simulation Results of the Frequency Domain Method Based on the PI Controller

Here, the frequency domain method-based PI controller is simulated. Adding the PI controller, the buck converter system's transfer function becomes:

$$G_o(s) = \frac{\tilde{y}(s)}{D(s)} = \frac{3{,}999{,}960 \times \left(K_P + \frac{K_I}{s}\right)}{S^2 + 803.333s + 135{,}998.4} \quad (34)$$

In this study, one of the requirements is that the PI controller is designed such that the closed-loop system achieves at least 75 degrees of phase margin. When the controller is tuned to achieve at least 75 degrees of phase margin (i.e., when $K_I = 1$, and $K_P = 0.23$), Fig. 13 shows that the gain margin is -0.151 dB, and the system achieved a stable loop. The step response of the system in Fig. 14 shows that the system has a fast response, a fast peak time, a small settling time, and a small overshoot. However, the steady-state error increased due to the low value of $K_P$.

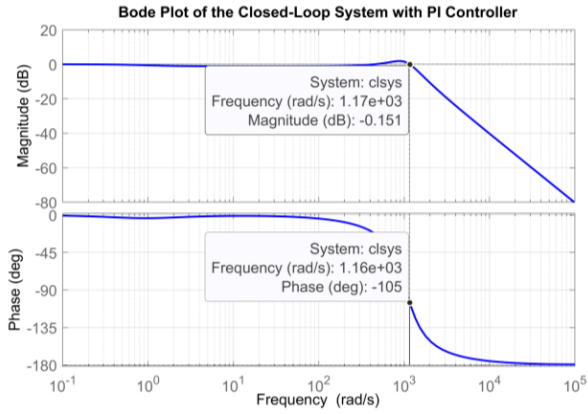

Fig. 13. Bode plot for when $K_I = 1$, and $K_P = 0.23$

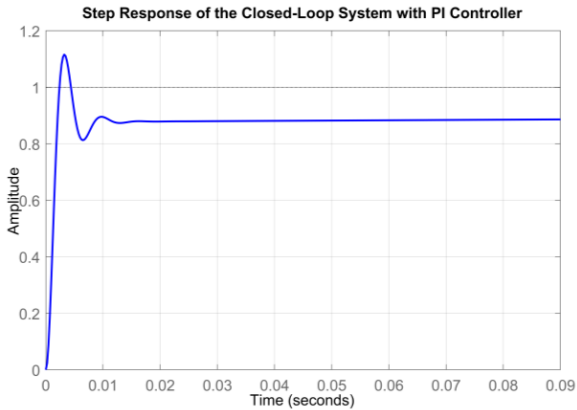

Fig. 14. Step response for when $K_I = 1$, and $K_P = 0.23$

When $K_I = 1$, and $K_P = 10$, Fig. 15 shows that the phase margin is 10 degrees, whereas the gain margin is 0.0428 dB. Also, the system achieved a stable loop. The step response of the system in Fig. 16 shows that the system's response is faster than in the previous scenario. Also, the steady-state error is reduced significantly. However, due to the increased value of $K_P$, the system experienced more overshoot and took more time to settle at a steady state. An excessive increase in $K_P$ can cause the phase margin to fall below -180 degrees. However, this will lead to system instability. Thus, an appropriate trade-off must be established to achieve the desired result.

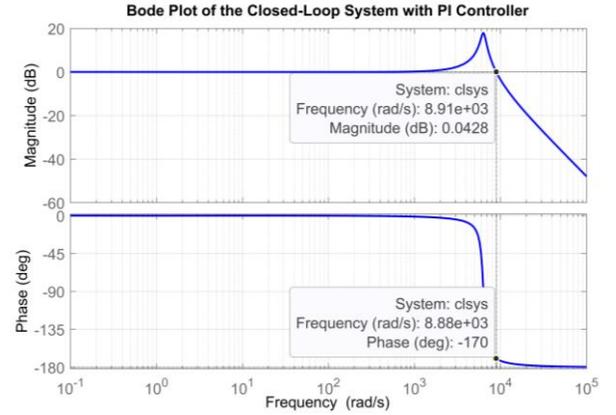

Fig. 15. Bode plot for when $K_I = 1$, and $K_P = 10$

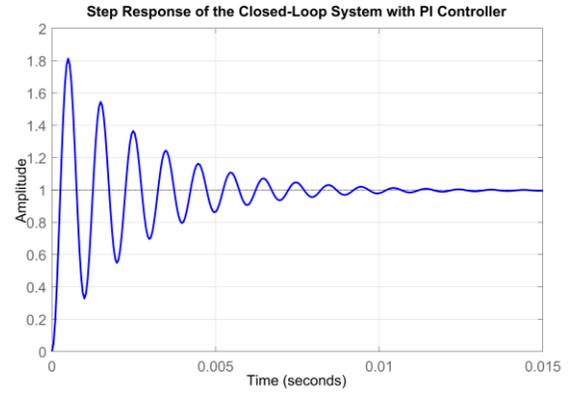

Fig. 16. Step response for when $K_I = 1$, and $K_P = 10$

### A. Complete Buck Converter Voltage Control Simulation Results

In this section, the complete model of the buck converter voltage control is presented, as shown in Fig. 17. Here, the reference voltage, $V_{ref}$ is set to 2 V, whereas the switching frequency, $f_s$ and the PWM sawtooth peak voltage, $V_s$ are set to 60 kHz and 10 V, respectively. The simulation results in Fig. 18 and Fig. 19 show that the output voltage and current remained stable despite variations in the load input voltage $V_g$.

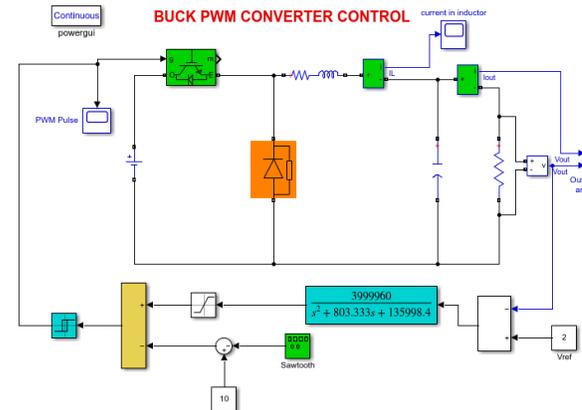

Fig. 17. The complete model of the buck converter voltage control

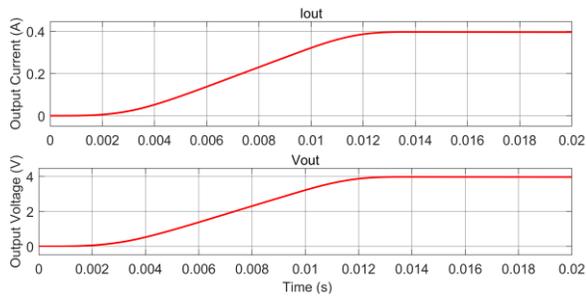
Fig. 18. Controlled buck converter output voltage and current at the input voltage, $v_g$ = 30 V, and R = 10

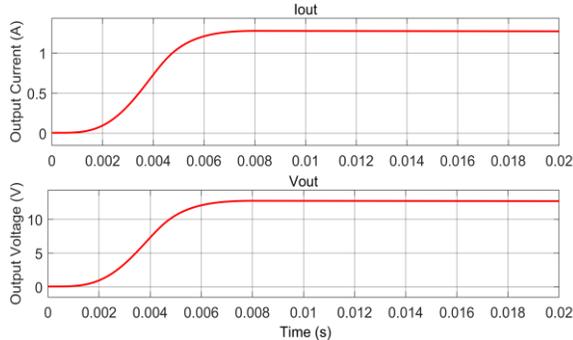
Fig. 19. Controlled buck converter output voltage and current at the input voltage, $v_g$ = 500 V and R = 10

## VII. CONCLUSION

This paper has demonstrated the modeling and control of a DC-DC buck converter using state-space averaging techniques to maintain a stable output voltage under varying input and load conditions. The converter was first mathematically modeled using a state-space average approach, enabling the derivation of a transfer function that relates the duty cycle to the output voltage. This model served as the foundation for designing and analyzing feedback control systems. A PI controller was designed using frequency-domain techniques. Simulation results revealed that the PI controller offers greater flexibility in tuning for specific performance requirements such as overshoot, settling time, and steady-state error. The complete simulation of the closed-loop buck converter confirmed its ability to sustain a constant output voltage despite significant variations in input voltage and load resistance. This underscores the effectiveness of combining state-space modeling with appropriate control strategies in power converter design.